\documentclass[aps,prl,twocolumn,superscriptaddress]{revtex4}

\usepackage{amsmath,amssymb}
\usepackage{graphicx}
\usepackage{times}

\begin{document}
\title{Relevance of the Heisenberg-Kitaev model for the honeycomb lattice iridates $A_2$IrO$_3$}

\author{Yogesh Singh}
\affiliation{Indian Institute of Science Education and Research Mohali, Sector 81, SAS Nagar, Manauli PO 140306, India}
\affiliation{I. Physikalisches Institut, Georg-August-Universit\"at G\"ottingen, D-37077, G\"ottingen, Germany}
\author{ S. Manni}
\affiliation{I. Physikalisches Institut, Georg-August-Universit\"at G\"ottingen, D-37077, G\"ottingen, Germany}
\author{J. Reuther}
\affiliation{Institut f\"ur Theorie der Kondensierten Materie, Karlsruhe Institute of Technology, D-76128 Karlsruhe, Germany}
\affiliation{Department of Physics and Astronomy, University of California, Irvine, CA 92697, USA}
\author{T. Berlijn}
\affiliation{Condensed Matter Physics and Materials Science Department, Brookhaven National Laboratory, Upton, New York 11973, USA}
\affiliation{Physics Department, State University of New York, Stony Brook, New York 11790, USA}
\author{R. Thomale}
\affiliation{Department of Physics, Stanford University, Stanford, CA 94305, USA}
\author{W. Ku}
\affiliation{Condensed Matter Physics and Materials Science Department, Brookhaven National Laboratory, Upton, New York 11973, USA}
\affiliation{Physics Department, State University of New York, Stony Brook, New York 11790, USA}
\author{S. Trebst}
\affiliation{Institute for Theoretical Physics, University of Cologne, 50937 Cologne, Germany}
\author{P. Gegenwart}
\affiliation{I. Physikalisches Institut, Georg-August-Universit\"at G\"ottingen, D-37077, G\"ottingen, Germany}
\date{\today}

\begin{abstract}
Combining thermodynamic measurements with theoretical density functional and thermodynamic calculations
we demonstrate that the honeycomb lattice iridates $A_2$IrO$_3$ ($A = $Na, Li) are magnetically ordered Mott insulators where the magnetism of the effective spin-orbital S=1/2 moments can be captured by a Heisenberg-Kitaev (HK) model with Heisenberg interactions beyond nearest-neighbor exchange.
Experimentally, we observe an increase of the Curie-Weiss temperature from $\theta \approx -125$~K for Na$_2$IrO$_3$ to $\theta \approx -33$~K for Li$_2$IrO$_3$, while the antiferromagnetic ordering temperature remains roughly the same $T_N \approx 15$~K for both materials.
Using finite-temperature functional renormalization group calculations we show that this evolution of $\theta$, $T_N$, the frustration parameter $f = \theta/T_N$, and the zig-zag magnetic ordering structure suggested for both materials by density functional theory can be captured within this extended HK model. 
Combining our experimental and theoretical results, we estimate that Na$_2$IrO$_3$ is deep in the magnetically ordered regime of the HK model ($\alpha \approx 0.25$), while Li$_2$IrO$_3$ appears to be close to a spin-liquid regime ($0.6 \leq \alpha \leq 0.7$).
\end{abstract}
\pacs{75.40.Cx, 75.50.Lk, 75.10.Jm, 75.40.Gb}

\maketitle

\paragraph{Introduction.--}
The fundamental importance of the Kitaev model, which describes the highly anisotropic exchange of SU(2) spin-1/2 moments on the honeycomb lattice, has quickly been appreciated for its rare combination of microscopic simplicity and an exact analytical solution \cite{Kitaev2006}. It has also become an archetypal example of a microscopic model that -- depending on the spatial anisotropy of its couplings -- harbors three of the currently most sought-after collective states in condensed matter physics \cite{Balents2010}: a gapless spin liquid with emergent Majorana fermion excitations, a gapped $Z_2$ spin liquid, and a topologically ordered phase with non-Abelian quasiparticle statistics (in the presence of a magnetic field perpendicular to the honeycomb lattice) \cite{Kitaev2006,Jiang2011}. Especially, physical realizations of topological states of the latter form which support Majorana fermion zero modes \cite{Stern2010} are intensely searched for in various candidate
systems including certain fractional quantum Hall systems \cite{MooreRead}, unconventional superconductors \cite{ReadGreen}, as well as heterostructures of topological insulators, semi-metals, or semiconductors with conventional $s$-wave superconductors \cite{FuKane2009,Sau2010,Alicea2010}, not only because of their fundamentally new character but also due to their possible application in topological quantum computation proposals \cite{Nayak2008}.

A direct realization of the Kitaev model could provide yet another alternative path to this goal. First proposals to engineer implementations of the Kitaev model were discussed in the context of optical lattices \cite{Duan2003} and superconducting circuits \cite{You2010}. More recently, it has been put forward that strong spin-orbit coupling in certain Mott insulating transition metal oxides \cite{Jackeli2009,Chaloupka2010} could inherently give rise to Kitaev-type couplings of effective spin-orbital degrees of freedom.  Among the best candidate materials are layered iridates of the form $A_2$IrO$_3$, which exhibit Mott insulating ground states and where the Ir$^{4+}$ form effective $S= 1/2$ moments as it was recently observed for Na$_2$IrO$_3$ \cite{Yogesh2010}.  On a microscopic level, it has been argued that the strong spin-orbit coupling in these 5d transition metal systems leads to orbital dependent anisotropic in-plane exchange that precisely mimics the Kitaev couplings
  \cite{Jackeli2009}. For real materials, however, further interactions will inevitably be present and in particular one might expect that the original spin exchange is not completely damped and isotropic Heisenberg interactions will compete with the anisotropic Kitaev couplings \cite{Chaloupka2010}.  
Such a Heisenberg-Kitaev (HK) model can be written down in its simplest form as 
\begin{equation}
H_{\rm HK} = (1-\alpha)\sum_{ij}\vec{\sigma_i}.\vec{\sigma_j}- 2\alpha\sum_{\gamma}\sigma^{\gamma}_i\sigma^{\gamma}_j \,,
\label{Eq:HK}
\end{equation}
where the $\sigma_i$ are Pauli matrices for the effective $S = 1/2$ and $\gamma = x, y, z$ labels the three different links for each spin of the honeycomb lattice. It has been shown \cite{Chaloupka2010} that the isotropic Heisenberg exchange in the first term of model~\eqref{Eq:HK} enters as antiferromagnetic coupling, while the anisotropic Kitaev exchange (in the second term) is ferromagnetic.
Varying the relative coupling $\alpha$ of the two exchange terms a sequence of three different phases has been found
\cite{Chaloupka2010,Jiang2011,Reuther2011}: a conventional  N\'eel antiferromagnet for $0 \leq \alpha \leq 0.4$, a so-called 
a stripy antiferromagnet for $0.4 \leq \alpha \leq 0.8$, and a spin-liquid state for $0.8 \leq \alpha \leq 1$.

But while the $A_2$IrO$_3$ materials have been appreciated from a theory perspective as possible candidate materials to look for Kitaev-like and HK-like physics \cite{Jackeli2009,Chaloupka2010,Reuther2011}, there has so far been very limited experimental data available for these layered iridates.  For Li$_2$IrO$_3$ there have been two conflicting reports \cite{Felner2002, Kobayashi2003}, with one report \cite{Felner2002} suggesting paramagnetic behavior between $T = 5$~K and 300~K without any sign of magnetic order,
while the second report \cite{Kobayashi2003} indicated an anomaly in the magnetic susceptibility below $T = 15$~K, which was also accompanied by a hysteresis between zero-field-cooled and field-cooled data, suggesting glassy behavior.  No heat capacity data has so far been available for Li$_2$IrO$_3$. 

For single crystal Na$_2$IrO$_3$ some of us have earlier shown \cite{Yogesh2010} Mott insulating behavior with antiferromagnetic ordering below $T_{\rm N} = 15$~K. Subsequent resonant magnetic x-ray scattering measurements \cite{Liu2011} on single crystals were consistent with either stripy or zig-zag magnetic order, with supplementary DFT calculations indicating that zig-zag order was the more likely magnetic ground state for Na$_2$IrO$_3$. 
This gave rise to another theoretical puzzle, since the original HK-model with nearest-neighbor exchange, i.e. model \eqref{Eq:HK}, allows for stripy magnetic ordering but not zig-zag order. Finite-temperature calculations \cite{Reuther2011} for model \eqref{Eq:HK} pointed to another discrepancy with experimental observations, since the theoretical calculations indicated
that the competition of the Heisenberg and Kitaev exchanges in model \eqref{Eq:HK} does not lead to a substantial suppression of the magnetic ordering transition with regard to the Curie-Weiss scale and the frustration parameter $f=|\Theta|/T_N$ was found to never exceed $f \approx 2$ \cite{Reuther2011}, while for Na$_2$IrO$_3$ experiments indicate $f \approx 8$ \cite{Yogesh2010}. 
Pieces of this puzzle were recently solved when it was shown that taking into account Heisenberg interactions beyond the nearest neighbor exchange can indeed stabilize the zig-zag ordering pattern \cite{Kimchi2011,RAT2011}. For antiferromagnetic exchanges, the latter are also expected to introduce geometric frustration. In the following we will expand this discussion of the role of further neighbor Heisenberg exchange and by providing a detailed comparison of theoretical and experimental results we will establish a microscopic description of the layered iridates $A_2$IrO$_3$ in terms of such an extended Heisenberg-Kitaev model.

Quickly summarizing our main results we report magnetic and heat capacity measurements on high quality polycrystalline samples of $A_2$IrO$_3$ ($A = $ Na, Li).  Magnetic measurements show local moment behavior with effective spin $S = 1/2$ moments.  Both magnetic and heat capacity measurements show sharp anomalies at $T_{\rm N} = 15$~K for both materials indicating bulk antiferromagnetic ordering.  For both materials our DFT calculations indicate that the most likely magnetic order is of zig-zag type. Finite-temperature functional renormalization group (FRG) calculations for an extended HK model including next-nearest ($J_2$) and next-to-next-nearest ($J_3$) neighbor antiferromagnetic Heisenberg  interactions are then used to confirm the type of magnetic order, and study the evolution of the Curie-Weiss temperature scale $\theta$, the ordering scale $T_{\rm N}$, and the frustration parameter $f = |\theta|/T_N$ as $\alpha$ and the various $J$'s are varied.  It must be emphasized that in contrast to the classical phase diagram of the extended Heisenberg-Kitaev model discussed earlier\cite{Kimchi2011} our FRG calculations are performed on the quantum level.
We show that the experimentally observed evolution of $\theta$, $T_N$, and $f$ and the observed magnetic order can all be very well captured within this extended HK model.  Comparison of experiments with calculations suggest that while the Kitaev term is small for Na$_2$IrO$_3$, the Li$_2$IrO$_3$ system with $0.6 \leq \alpha \leq 0.7$ sits quite close to the spin-liquid state in the Kitaev limit $\alpha \geq 0.8$.

\begin{figure}[t]
\includegraphics[width=2.35in,angle=-90]{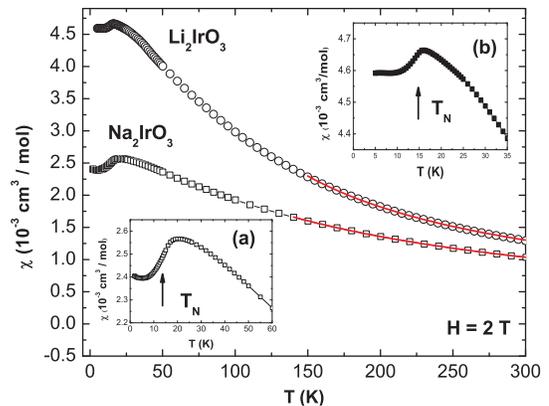}
\caption{(Color online) Magnetic susceptibility $\chi$ versus temperature $T$ for $A_2$IrO$_3$ ($A =$~Na, Li).  The fit by the Curie-Weiss (CW) expression $\chi = \chi_0 + C/(T-\theta)$ is shown as the curve through the data.  The insets (a) and (b) shows the anomaly at the antiferromagnetic ordering for the Na and Li systems, respectively.   
\label{Fig-chi}}
\end{figure}

\paragraph{Magnetic Susceptibility.--} The magnetic susceptibility $\chi = M/H$ versus $T$ data for $A_2$IrO$_3$ ($A = $ Na, Li) are shown in Fig.~\ref{Fig-chi}.  The $\chi(T)$ data between $T = 150$~K and 300~K were fit by the Curie-Weiss expression $\chi = \chi_0 + {C\over T-\theta}$ with $\chi_0$, $C$, and $\theta$ as fitting parameters.  The fit, shown in Fig.~\ref{Fig-chi} as the solid curve through the data, gives the values $\chi_0 = 3.6(4)\times10^{-5}$~cm$^3$/mol, $C = 0.40(2)$~cm$^3$~K/mol, and $\theta = - 125(6)$~K, for Na$_2$IrO$_3$ and, $\chi_0 = 8.1(7)\times10^{-5}$~cm$^3$/mol, $C = 0.42(3)$~cm$^3$~K/mol, and $\theta = - 33(3)$~K, for Li$_2$IrO$_3$, respectively.  Assuming a $g$-factor $g = 2$ the above values of $C$ correspond to an effective moment of $\mu_{\rm eff} = 1.79(2)~\mu_{\rm B}$ and $\mu_{\rm eff} = 1.83(5)~\mu_{\rm B}$, for Na$_2$IrO$_3$ and Li$_2$IrO$_3$, respectively.  These values of $\mu_{\rm eff}$ are close to the value 1.74~$\mu_{\rm B}$ expected for spin~=~1/2 moments.  
This local-moment formation along with the insulating resistivity (see auxiliary material \cite{Aux}) indicates that like its sister compound Li$_2$IrO$_3$ is indeed a Mott insulator.
The value of the Weiss temperature $\theta = -33(3)$~K for Li$_2$IrO$_3$ further suggests that the effective antiferromagnetic exchange interactions have weakened when compared to the Na$_2$IrO$_3$ system. However, the $\chi(T)$ data for Li$_2$IrO$_3$ also show an anomaly at $15$~K suggesting that an antiferromagnetic transition occurs at roughly the same temperature as for Na$_2$IrO$_3$.  This is further supported by our heat capacity results presented below.  The insets (a) and (b) in Fig.~\ref{Fig-chi} show the $\chi(T)$ data %for Na$_2$IrO$_3$ and Li$_2$IrO$_3$, respectively, 
at low temperatures to highlight the anomaly seen at the onset of the antiferromagnetic transition below $T_{\rm N} = 15$~K in both materials.    

\begin{figure}[t]
\includegraphics[width=2.35in, angle=-90]{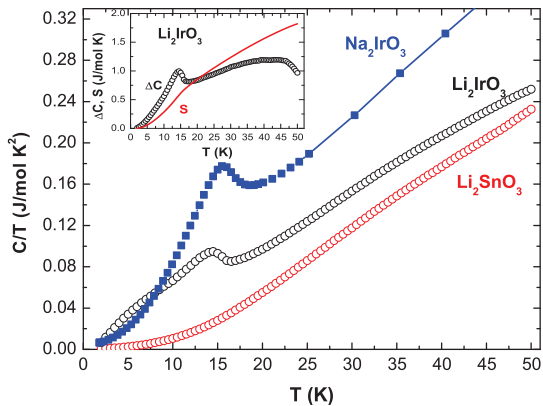}
\caption{(Color online) Heat capacity divided by temperature $C/T$ versus $T$ data between $T = 1.8$~K and 50~K for $A_2$IrO$_3$ ($A = $Na, Li) and the non-magnetic analog Li$_2$SnO$_3$.  The anomaly at $T_{\rm N} = 15$~K for both $A_2$IrO$_3$ materials indicates onset of bulk antiferromagnetic order.  The inset shows the difference heat capacity $\Delta C(T)$ and the difference entropy $S(T)$ for Li$_2$IrO$_3$.  
\label{Fig-heatcap}}
\end{figure}

\paragraph{Heat Capacity.--} In Fig.~\ref{Fig-heatcap} we show heat capacity data divided by temperature $C/T$ versus temperature $T$ for $A_2$IrO$_3$ ($A =$~Na, Li), and for the non-magnetic analog Li$_2$SnO$_3$.  The anomaly at $T_{\rm N} = 15$~K in the data for both $A_2$IrO$_3$ ($A =$~Na, Li) materials confirms bulk magnetic ordering.  A small bump is also observed around $T = 5$~K in the $C(T)$ for Li$_2$IrO$_3$.  This most likely arises due to a small amount ($\leq 5\%$) of disorder in the sample.\cite{Soham}  
The magnetic contribution $\Delta C(T)$ for Li$_2$IrO$_3$ shown in the inset of Fig.~\ref{Fig-heatcap} was obtained by subtracting the $C(T)$ data of Li$_2$SnO$_3$ from the data of Li$_2$IrO$_3$. The latter reveals a clearly more visible 
 lambda-like anomaly at $T_{\rm N} = 15$~K. A slight depression of $T_{\rm N}$ in an applied magnetic field of $H = 9$~T was observed (not shown) which points to the antiferromagnetic nature of the magnetic ordering in Li$_2$IrO$_3$.  The entropy $S(T)$ obtained by integrating the $\Delta C/T$ versus $T$ data is also shown in Fig.~\ref{Fig-heatcap} inset.  Just above $T_{\rm N}$ the entropy is only about 15\%Rln2.  Such a reduced entropy at the transition was also observed earlier for single crystalline Na$_2$IrO$_3$.\cite{Yogesh2010}  The small entropy points to the reduced ordered moment and the possible proximity to a non-magnetic ground state.

\paragraph{Magnetic Ordering.--} 
From the similarities in the anomalies seen in $\chi$ and $C$ data for both the Na and Li systems, it would seem likely that the kind of magnetic order would also be similar for the two. 
To resolve the magnetic structure for Li$_2$IrO$_3$, spin density function calculations within the LDA+U+SO approximation were performed for the N\'eel, stripy, and zig-zag configurations with the moments constrained along the crystallographic axes ~\cite{Kobayashi2003,Aux}. 
The results are summarized in Table~\ref{tab:tab1}. We find that as for Na$_2$IrO$_3$~\cite{Liu2011}, the zig-zag configuration has the lowest energy and is hence the most likely magnetic structure for Li$_2$IrO$_3$.

\begin{table}[b]
\begin{ruledtabular}
\begin{tabular}{ccccc}
E$_{\rm{tot}}$per Ir(meV) & zig-zag  & stripy & N\'eel \\
\hline 
Li$_2$IrO$_3$	& 0	& 24	& 18	 \\
\end{tabular}
\end{ruledtabular}
\caption{ 
Total energy E$_{\rm{tot}}$ per Ir for Li$_2$IrO$_3$ for three antiferromagnetic configurations, obtained from collinearly constrained LDA+U+SO simulations. 
}\label{tab:tab1}   
\end{table}

We now turn to the evolution of magnetic properties as we go from the Na to  the Li compound.  From our $\chi(T)$ data above we find that the Curie-Weiss temperature $\theta$ decreases from $\approx -120$~K to $\approx -33$~K on going from Na$_2$IrO$_3$ to Li$_2$IrO$_3$ possibly indicating that the effective magnetic interactions are weaker for Li$_2$IrO$_3$.  Surprisingly however, both $\chi(T)$ and $C(T)$ show that both materials order magnetically at roughly the same temperature $T_N \approx 15$~K\@.  The frustration parameter $f = \theta/T_N$, however, reduces from $\approx 8$ for Na$_2$IrO$_3$ to $\approx 2$ for Li$_2$IrO$_3$.  

In previous theoretical calculations \cite{Reuther2011} for the thermodynamics of the HK model \eqref{Eq:HK}, the ordering temperature $T_N$ was found to be largely insensitive to variations of $\alpha$ whereas $\theta$ was found to decrease monotonically with $\alpha$ within the stripy magnetic phase.   The increase of the Weiss temperature scale on increasing $\alpha$ is a direct consequence of the fact that the two coupling terms in the HK model \eqref{Eq:HK} enter with opposite coupling signs. Increasing the relative strength of the ferromagnetic Kitaev term thus leads to an increase of the  Weiss temperature scale. These theoretical trends thus seem to match well with what is observed in our experiments.

There are, however, two issues where experiments differ from predictions for the HK model.  First, the zig-zag magnetic order obtained from DFT for both Na$_2$IrO$_3$ and Li$_2$IrO$_3$ is not one of the three phases of the HK model \cite{Chaloupka2010}.  Secondly, a maximum frustration parameter $f \approx 2$ was found in calculations for the HK model\cite{Reuther2011}, which is much smaller than the experimentally observed value $f \approx 8$ for Na$_2$IrO$_3$\cite{Yogesh2010}.  
%These discrepancies pose doubts on the applicability of the HK model in describing the $A_2$IrO$_3$ ($A =$ Na, Li) materials. 

To resolve these discrepancies it has recently been argued \cite{Kimchi2011} that further nearest neighbor antiferromagnetic Heisenberg exchange interactions should be added to the original HK model, which can indeed stabilize the zig-zag magnetic order. It was further demonstrated \cite{Kimchi2011} that the experimental magnetic susceptibility data for $A_2$IrO$_3$ ($A =$ Na, Li) materials can only be fit when including the Kitaev term in this expanded microscopic model.

\begin{figure}[t]
\includegraphics[width=2.25in,angle=-90]{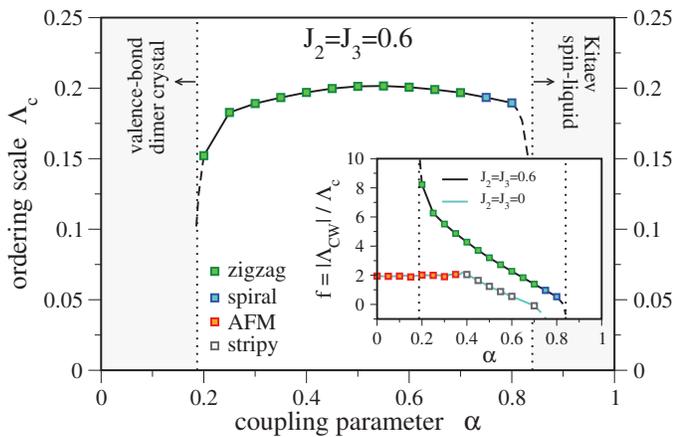}
\caption{(Color online) Ordering scale $\Lambda_c$ obtained from the FRG calculations of the HK-$J_2,J_3$ model as a function of $\alpha$ for $J_2, J_3 = 0.6$. The dashed line indicates the crossover between regions with magnetic order and regions with no long range order at both low and high $\alpha$, respectively. A regime of enhanced numerical uncertainties is seen near $\alpha \approx$ 0.8. The inset shows the frustration parameter f as a function of $\alpha$ for $J_2$, $J_3=0.6$ in comparison to the pure Heisenberg-Kitav model with $J_2$, $J_3=0$.  
\label{Fig-T-vs-alpha}}
\end{figure}

The inclusion of further than nearest-neighbor antiferromagnetic interactions is further expected to introduce geometric frustration in addition to the frustration arising from the competition between the Heisenberg and Kitaev couplings of the original model.  We have therefore expanded our  FRG calculations \cite{Reuther2011} to such an extended HK-$J_2$,$J_3$ model and determine its thermodynamic properties by extracting the high-temperature CW behavior (from the RG flow), the onset of magnetic ordering (from the breakdown of the RG flow), and the nature of the various ground states by calculating 
momentum-resolved magnetic susceptibility profiles as further detailed in  the auxiliary material \cite{Aux}.
We focused our calculations on the parameter regime $0.2 \leq \alpha \leq 0.8$ and $0 \leq J_2, J_3 \leq 1$.  
A representative plot of the ordering scale $\Lambda_c$ as a function of $\alpha$ for fixed $J_2 = J_3 = 0.6$ is shown in Figure~\ref{Fig-T-vs-alpha} with the inset showing the evolution of the frustration parameter $f$. 
Our calculations indicate that zig-zag order is stabilized for an extended range $0.25 \leq \alpha \leq 0.7$ in agreement with recent calculations for the HK-$J_2,J_3$ model \cite{Kimchi2011}. 
Around the Kitaev limit for $\alpha \geq 0.8$ we find an extended non-magnetic spin-liquid phase, which connects directly to the one of the original HK model \eqref{Eq:HK}.
For $\alpha \leq 0.2$ we obtain another non-magnetic ground state, evidently arising from the further nearest-neighbor exchange $J_2,J_3$ frustrating the nearest-neighbor $J$ and suppressing the N\'eel state in favor of a valence bond dimer crystal \cite{RAT2011}.  

Our results summarized in Fig.~\ref{Fig-T-vs-alpha} further indicate that the potentially counter-intuitive experimental observation of $T_N$ staying roughly the same  in going from Na$_2$IrO$_3$ to Li$_2$IrO$_3$ even though $\theta$ decreases by a factor of $\approx 4$ in fact agrees well with our calculations showing that the ordering scale stays more or less constant for the zig-zag ordered ground state in the regime $0.25 \leq \alpha \leq 0.7$.  Finally, we note that our calculations also indicate that the frustration parameter $f = \theta/T_N$ decreases monotonically with $\alpha$ in the region where magnetic order is found (see the inset in Fig.~\ref{Fig-T-vs-alpha}), which is a direct consequence of the Curie-Weiss scale $\theta$ decreasing monotonically in this region.  Interestingly, for small $\alpha$ the geometric frustration induced by the further nearest-neighbor exchange becomes more evident and the parameter $f$ reaches values much larger than obtained for the original HK model \cite{Reuther2011} and in fact becomes comparable in size to what is observed experimentally for Na$_2$IrO$_3$ where $f \approx 8$.  

To place the $A_2$IrO$_3$ materials on the diagram in Fig.~\ref{Fig-T-vs-alpha} we note that the zig-zag ordered ground state indicated in DFT calculations for both materials is found only in the parameter range  $0.25 \leq \alpha \leq 0.7$ in the presence of significant second and third neighbor exchange.  Additionally, an enhanced frustration parameter $f$ is found only for small $\alpha \geq 0.25$ before the system transitions to a non-magnetic ground state (for $\alpha \leq 0.2$) and the ordering temperature starts to drop drastically.  We therefore place Na$_2$IrO$_3$ at $\alpha \sim 0.25$. In contrast, Li$_2$IrO$_3$ with zig-zag order at $T_N = 15$~K and $f \approx 2$ can be placed at $\alpha \geq 0.65$ putting it considerably closer to the spin-liquid regime \cite{FootnoteSusceptibility} around the Kitaev limit beyond $\alpha \geq 0.8$.

Finally, we note that in going from the Na to the Li system the $a, b$ lattice parameters are reduced by $\approx 4.5\%$ while the $c$ parameter is reduced by $\approx 10\%$.  Thus, substituting Na by Li is equivalent to preferentially applying chemical pressure along the $c$ axis ($\perp$ to the honeycomb planes).  This most likely leads to the IrO$_6$ octahedra becoming more symmetrical within the $ab$-plane which in turn enhances the parameters $\eta_{1,2}$ (in the notation of Ref.~\onlinecite{Chaloupka2010}) leading to an increased Kitaev coupling.  This is consistent with the value of $0.6 \leq \alpha \leq 0.8$ estimated above for Li$_2$IrO$_3$ which puts its closer to the Kitaev limit.  
   
In summary, we have shown that magnetic properties of the Mott insulating iridates $A_2$IrO$_3$, 
in particular the evolution of  thermodynamic observables, i.e. the Curie-Weiss temperature $\theta$ and ordering temperature $T_N$, as well as their low-temperature magnetic order can be captured within an extended Heisenberg-Kitaev model. 
Our detailed comparison of experiment and theory in particular suggests that while Na$_2$IrO$_3$ is located deep in a magnetically ordered regime, Li$_2$IrO$_3$ lies close to the spin-liquid regime around the Kitaev limit ($\alpha \geq 0.8$). 
Future experiments will further investigate whether the application of $c$-axis pressure can push these systems deeper into the orbitally dominated regime, and in particular whether Li$_2$IrO$_3$ can be pushed into the spin-liquid phase.

\paragraph{Acknowledgments.--} YS acknowledges support from Alexander von Humboldt foundation and from DST, India.  SM acknowledges support from the Erasmus Mundus Eurindia Project.  JR is supported by the Deutsche Akademie der Naturforscher Leopoldina through grant LPDS 2011-14.  RT is supported by an SITP fellowship by Stanford University.  WK and TB acknowledge support from DOE \# DE-AC02-98CH10886

\clearpage

\appendix

\begin{widetext}
%\vskip 12mm

\begin{center}
\large\bf  Auxiliary Material
\end{center}

\section*{Synthesis and Structure}
\paragraph*{Synthesis.--}
Polycrystalline samples of $A_2$IrO$_3$ ($A = $ Na, Li) and Li$_2$SnO$_3$ were synthesized by solid state synthesis.  High purity starting materials $A_2$CO$_3$ ($A = $ Na, Li) (99.995\% Alfa Aesar) and Ir metal powder ($\geq 99.95\%$ Alfa Aesar) or SnO$_2$ (99.995\% Alfa Aesar) were mixed in the ratio 1.05 : 1 and placed in an alumina crucible with a lid and given heat treatments between 750~$^{\circ}$C and 950~$^{\circ}$C in steps of 50~$^{\circ}$C with intermediate grindings and pelletizing after each step.  The Li$_2$SnO$_3$ sample was given a further heat treatment at 1000~$^{\circ}$C\@.  In an attempt to grow single crystals the Li$_2$IrO$_3$ sample was dissolved in excess LiCl flux at 850~$^{\circ}$C for 6~hrs and then cooled to 590~$^{\circ}$C at 3~$^{\circ}$C/hr.  After washing majority of the LiCl flux with de-ionized water we obtained a fine black powder which turned out to be highly ordered Li$_2$IrO$_3$ polycrystalline samples.  This powder was pelletized and sin
 tered at 900~$^{\circ}$C for 48~hrs to get a hard pellet for resistivity and heat capacity measurements.
  
\begin{figure}[b]
\includegraphics[width=.6\columnwidth,angle=-90]{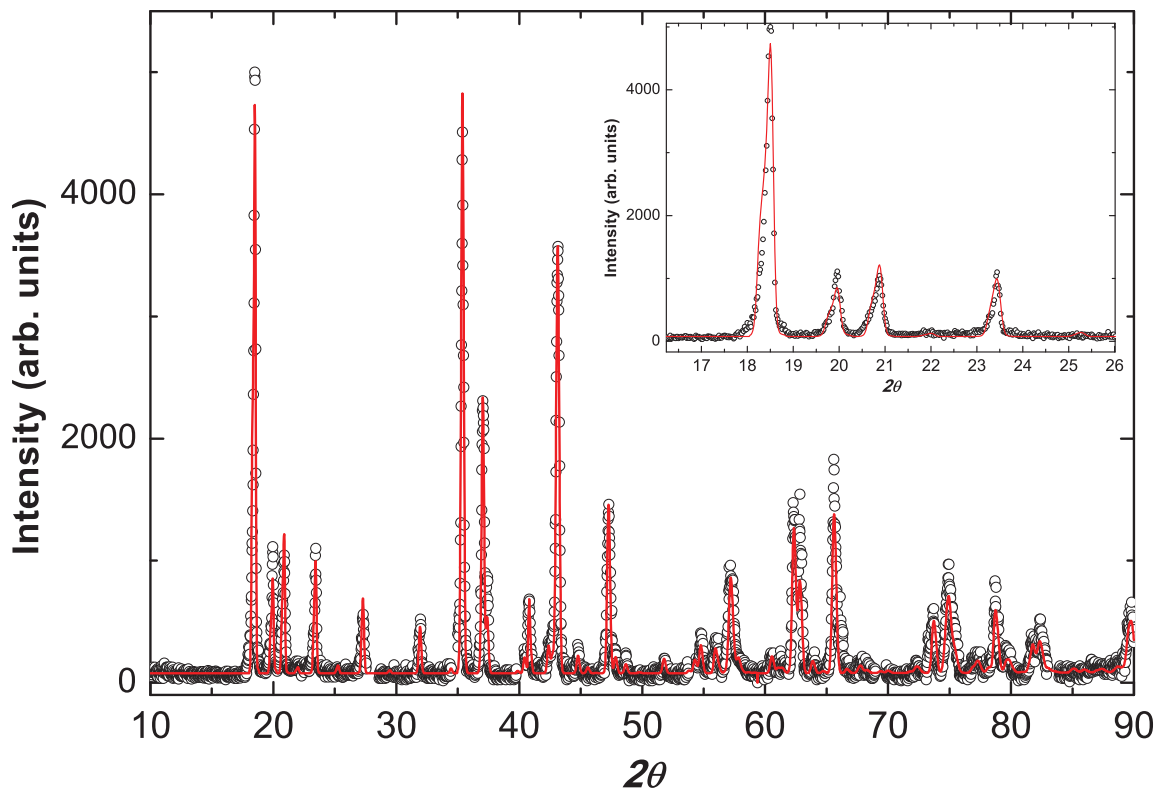}
\caption{Rietveld refinement of the X-ray diffraction data of polycrystalline Li$_2$IrO$_3$.  The closed symbols represent the observed data, the solid lines represent the fitted pattern.  Regions where small ($\leq 10\%$) peaks of LiCl appeared have been excluded in the fits.  The inset shows an expanded view of the data at low angles.   
\label{Fig-xrd}}
\end{figure}

\paragraph*{Structure.--} Powder x-ray diffraction (XRD) scans of polycrystalline $A_2$IrO$_3$ ($A =$ Na, Li) and Li$_2$SnO$_3$ were found to be single phase and could be indexed to the monoclinic \emph{C2/c} (No.~15) structure.  Rietveld refinements,\cite{Rietveld} of the x-ray patterns gave the unit cell parameters $a$~=~~5.4187(2) \AA , $b$~=~9.3689(3) \AA\, $c$~=~10.7731(5) \AA\, and $\beta = 99.598(18)~^{\circ}$ for Na$_2$IrO$_3$ and $a$~=~~5.1678(4) \AA , $b$~=~8.9347(7) \AA\, $c$~=~9.7825(4) \AA\, and $\beta = 99.998(14)~^{\circ}$ for Li$_2$IrO$_3$.  A representative XRD scan for Li$_2$IrO$_3$ obtained by LiCl flux method is shown in Fig.~\ref{Fig-xrd} along with the Rietveld refinement shown as the curve through the data.  Regions where small peaks of LiCl appeared have been excluded in the fits.  The inset of Fig.~\ref{Fig-xrd}  shows the XRD data and Rietveld refinement at low angles on an expanded scale.  The highly ordered nature of the sample can be seen from the
  separation of 
 the lines between $2\theta = 19$ and $22~^{\circ}$ which in a disordered sample would appear merged together and would also show a highly anisotropic shape with a tail extending to the higher angle side of the peak.\cite{Malley2008,Soham}

\section*{Resistivity} The resistivity $\rho$ versus temperature $T$ for Li$_2$IrO$_3$ is shown in Fig.~\ref{Fig-RES}.  The data clearly point to insulating behavior with a room temperature value $\approx 35~\Omega$~cm and an activation gap $\approx 700$~K\@.  The insulating resistivity coupled with the local-moment magnetism indicates that Li$_2$IrO$_3$ is a Mott insulator.

\begin{figure}[h]
\includegraphics[width=\columnwidth]{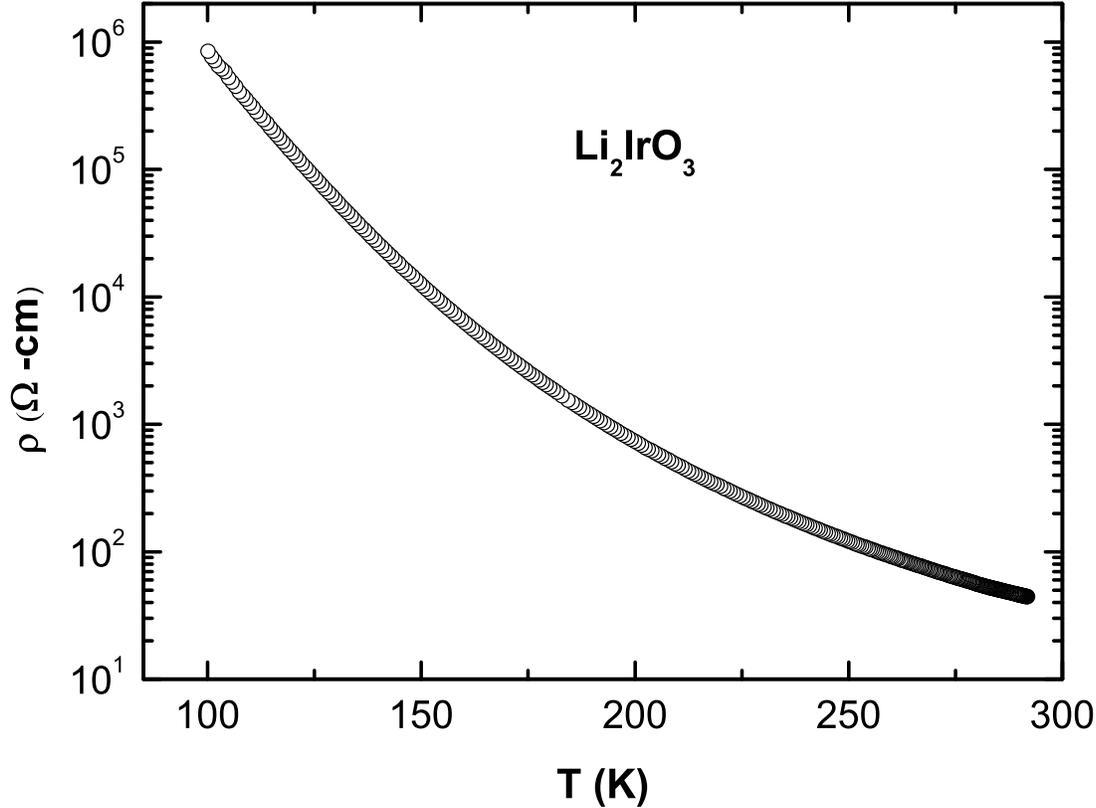}
\caption{Resistivity $\rho$ versus temperature $T$ data for polycrystalline Li$_2$IrO$_3$ shown on a semi-log plot.   
\label{Fig-RES}}
\end{figure}

\begin{figure}[t]
\includegraphics[width=0.8\columnwidth,angle=-90]{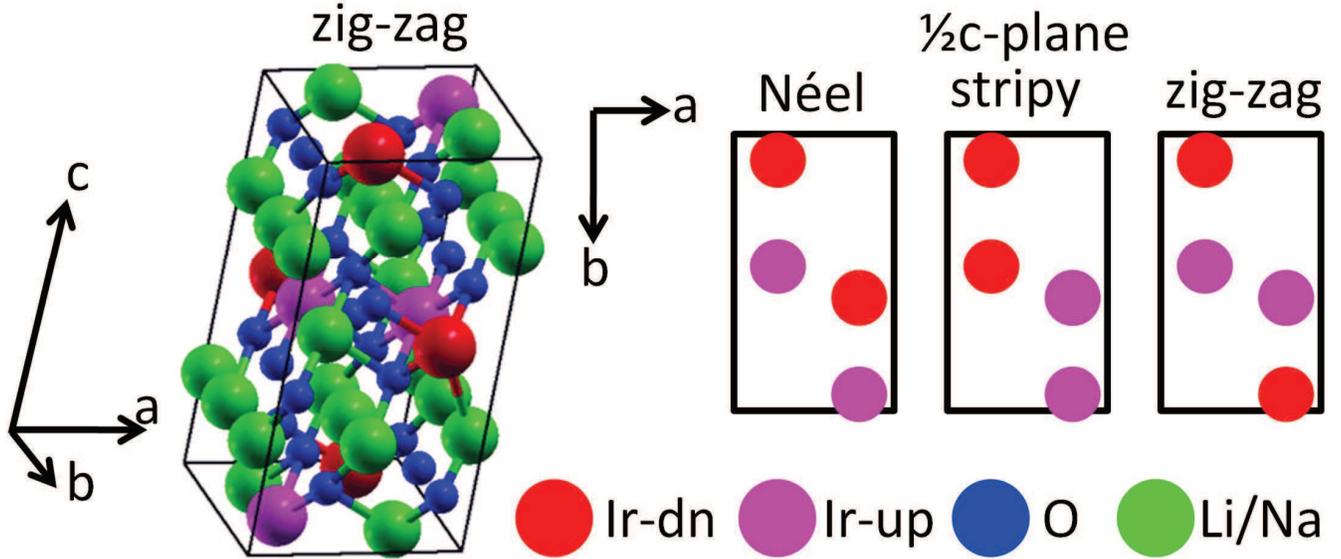}
\caption{\label{fig:figsup1}
(Up) Total energy per Ir in meV obtained from LDA+U+SO simulations of three antiferromagnetic configurations with the moments constrained along the three crystallographic axes. (Down) Lattice and magnetic structure.}
\end{figure}

\begin{figure}[t]
\includegraphics[width=.75\columnwidth,angle=-90]{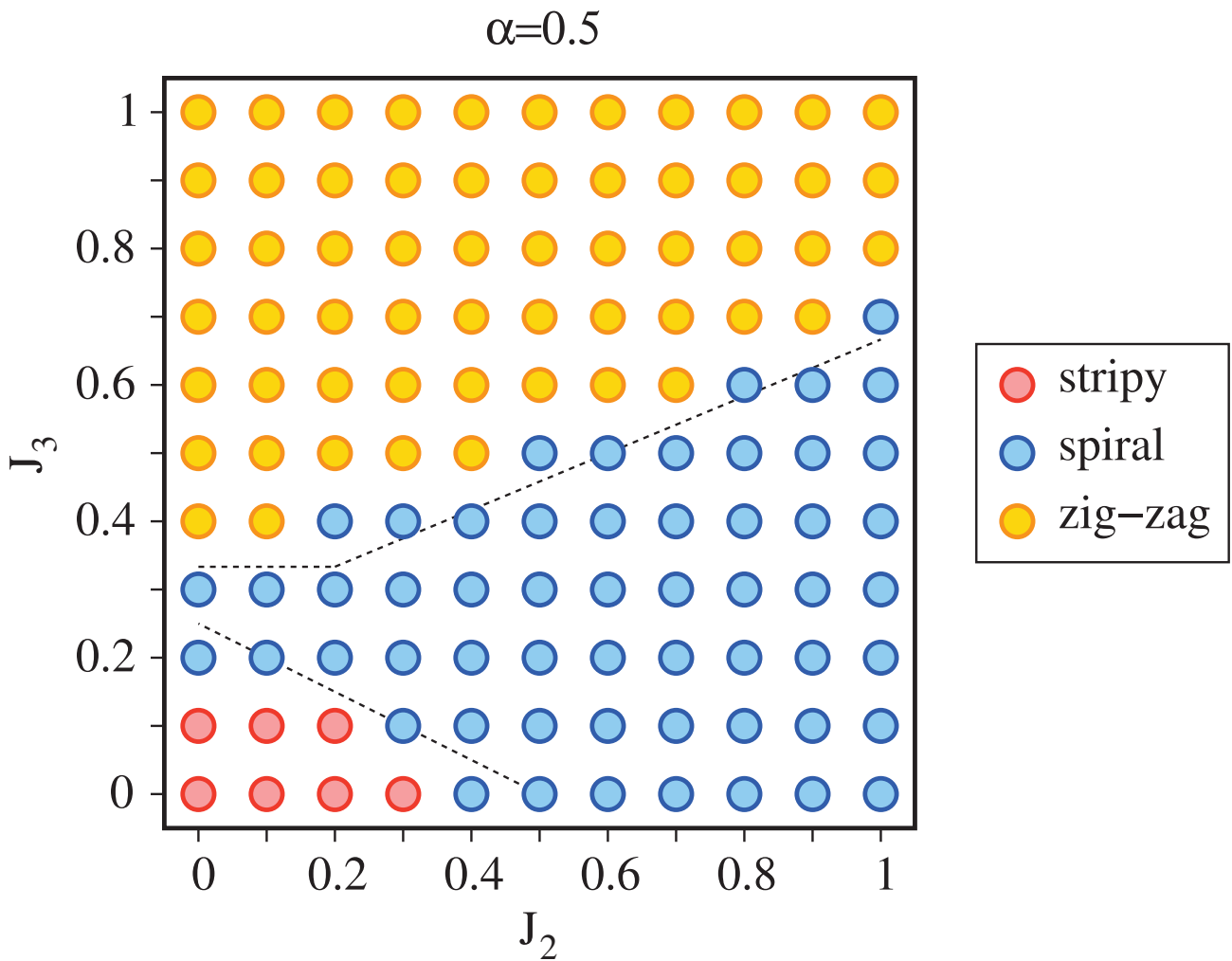}
\caption{Phase diagram of the extended Heisenberg-Kitaev model \eqref{Eq:NewHamiltonian} for $\alpha=0.5$ with $0\leq J_2, J_3\leq1$. Here, the $J_2$-$J_3$-parameter space is scanned in steps of 0.1. The dotted lines illustrate the classical phase boundaries obtained in Ref.~\onlinecite{Kimchi2011}.}
\label{Fig-phases}
\end{figure}

\begin{figure*}[t]
\includegraphics[width=0.35\linewidth,angle=-90]{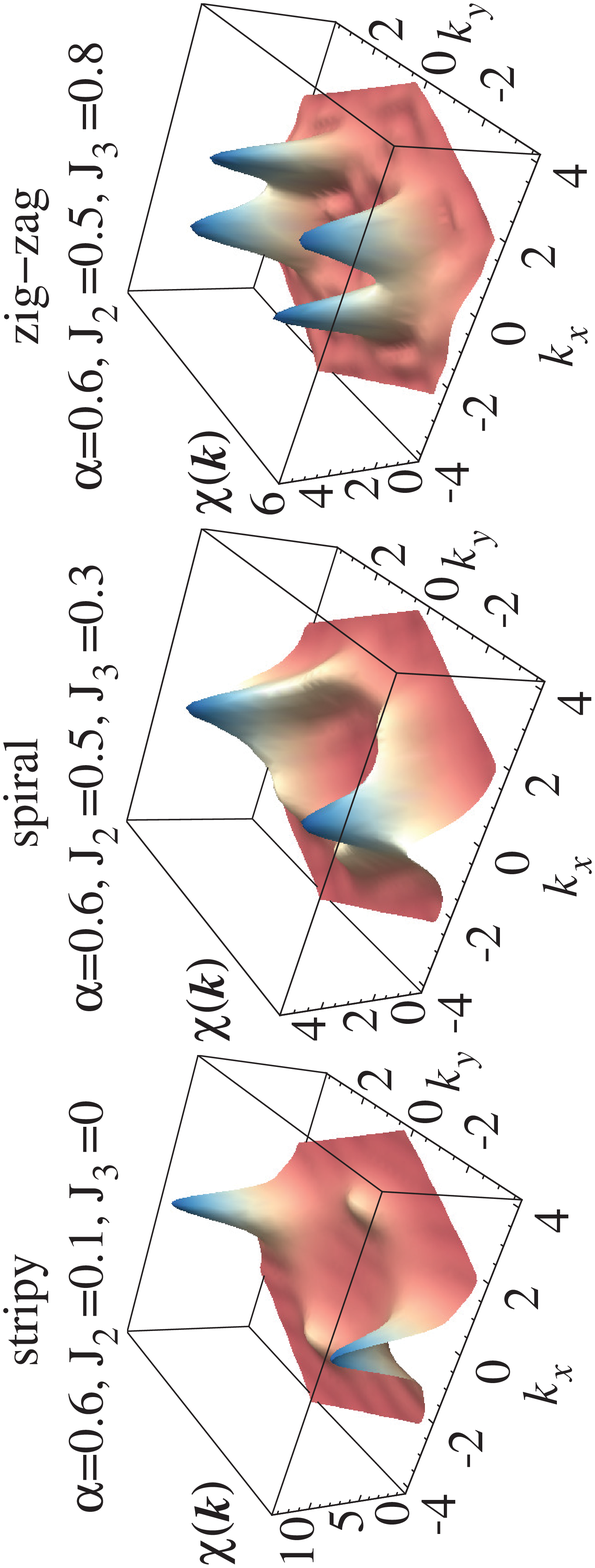}
\caption{Static magnetic spin susceptibility plotted in the entire extended (second) Brillouin zone. The three pictures represent the characteristic susceptibility profiles in the case of stripy, spiral and zig-zag order.
\label{Fig-order}}
\end{figure*}

\section*{Details of the Density Functional Theory Calculations}
We applied the WIEN2K~\cite{blaha} implementation of the all electron linearized augmented plane wave method in the LDA+U+SO approximation, within the second variational treatment using U=3eV and J=0.6eV. For the lattice structure the empirical monoclinic lattice structure C2/c ~\cite{kobayashi} was used.  The moments directions were constrained along the three crystallographic axes. The moments connected by half a unit cell in the c direction, were aligned antiferromagnetically. The basis size was determined by RKmax=8.0 and the Brillouin zone was sampled with a 5x5x2-mesh. For the zig-zag, N\'eel and stripy configuration, the moment direction with the lowest total energy is found to be `a', `c' and `a' respectively.  We note that the constrained moment directions investigated in this study are not meant to represent the moment direction of the theoretical ground state, which in general could be away from the crystallographic axes and non-collinear. We also note that another
  closely related space group C2/m has been suggested for Li$_2$IrO$_3$ in the literature~\cite{Malley2008}. Since our X-ray diffraction patterns could be fit equally well with either space group, we have used C2/c in our DFT calculations for Li$_2$IrO$_3$ to be consistent with what was previously used in calculations for Na$_2$IrO$_3$~\cite{liu}.

\section*{Functional Renormalization Group Calculations}
To capture the experimental data we consider an extended Heisenberg-Kitaev model which is augmented by second and third nearest neighbor Heisenberg couplings $J_2$ and $J_3$, respectively. Explicitly, the Hamiltonian is given by
 
\begin{eqnarray}
  H_{\rm HK-J_2J_3}& = &
    (1-\alpha)\left(\sum_{\langle ij \rangle} + J_2 {\sum_{\langle\langle i,j \rangle\rangle}} + J_3 {\sum_{\langle\langle\langle i,j \rangle\rangle\rangle}}\right ) \vec{\sigma}_i\cdot \vec{\sigma}_j 
     \nonumber\\
    &&- 2\alpha\sum_{\gamma}\sigma^{\gamma}_i\sigma^{\gamma}_j,
   \label{Eq:NewHamiltonian}
\end{eqnarray}
where nearest-neighbor exchange is indicated by the $\langle . \rangle$ brackets, and second and third nearest
neighbor exchange by $\langle\langle . \rangle\rangle$ and $\langle\langle\langle . \rangle\rangle\rangle$,
respectively. $J_2$ and $J_3$ are given in units of the nearest neighbor Heisenberg coupling. 

Using a pseudo-fermion functional renormalization group (FRG) calculations we were able to
(i) determine the phase diagram, 
(ii) characterize the magnetic ordering of the different phases, 
and (iii) calculate thermodynamic properties such as the Curie-Weiss scale, the ordering scale, and the frustration parameter. 
Quickly summarized, the main outcome of the FRG calculation is the magnetic spin susceptibility as the linear response to a small external magnetic field, calculated via Kubos formula. Note that the magnetic field always points along one of the cubic axes. Since the FRG naturally computes the spin susceptibility as a function of a frequency cutoff parameter $\Lambda$, we are able to extract thermodynamic properties of the model system via an identification of precisely this frequency cutoff parameter $\Lambda$ with a temperature $T$.
Our calculations use cluster sizes of the hexagonal lattice of 112 lattice sites.
For further technical details about the pseudo-fermion FRG method and the extraction of the Curie-Weiss scales and ordering scales from the FRG data we refer to Refs.~\onlinecite{Reuther2010,Reuther2011}. 

\begin{figure}[t]
\includegraphics[width=0.85\linewidth,angle=-90]{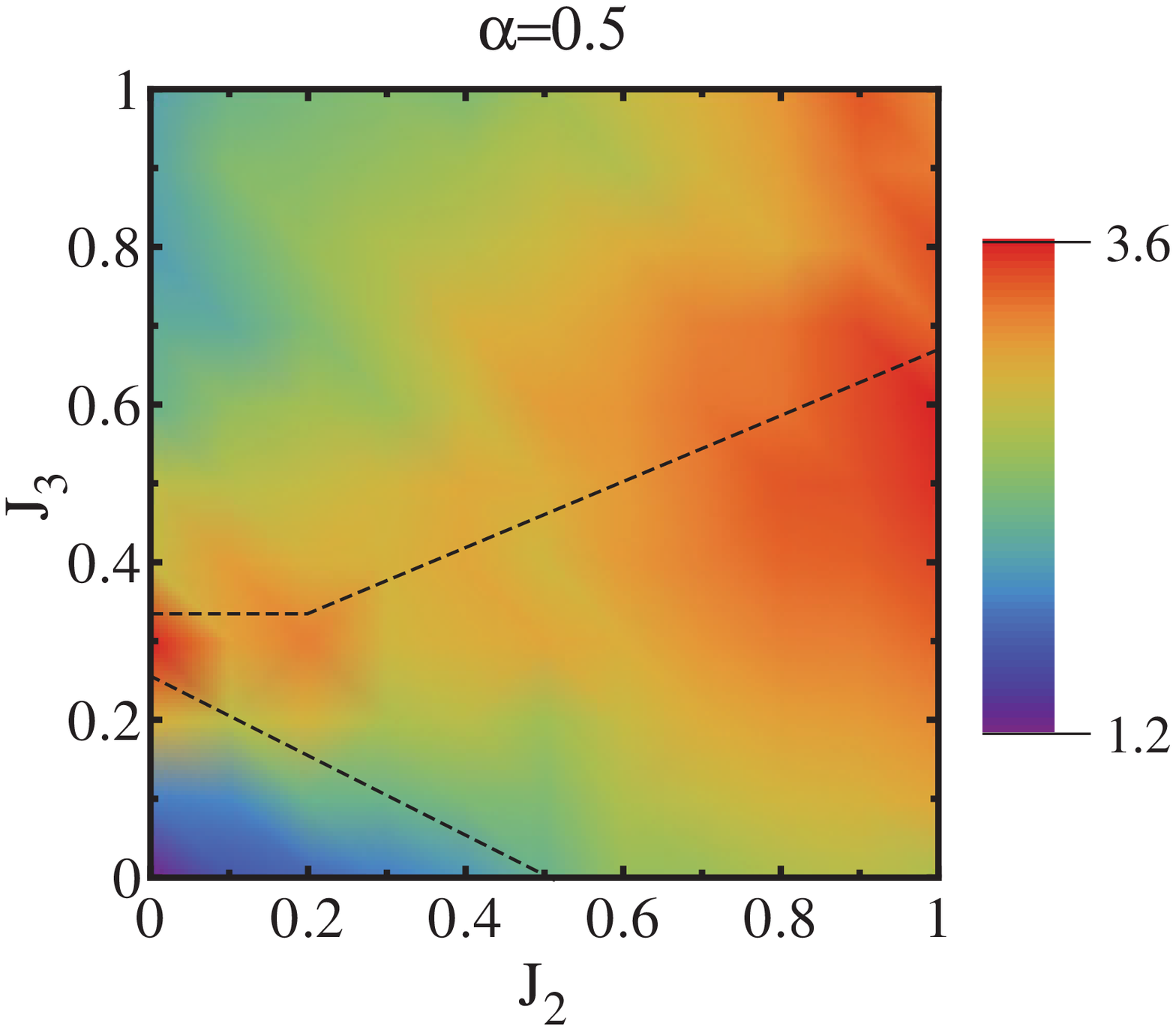}
\caption{Total color profile plot of the frustration parameter $f=\theta/T_N$ for $\alpha=0.5$ as a function of $J_2$ and $J_3$. The dotted lines illustrate the classical phase boundaries obtained in Ref.~\onlinecite{Kimchi2011}.
\label{Fig-f}}
\end{figure}

\paragraph*{Phase diagram.--}
First, we turn to the overall phase diagram of model \eqref{Eq:NewHamiltonian}. This phase diagram has recently been calculated \cite{Kimchi2011} on a classical level, which revealed four phases with different types of long-range order, i.e., antiferromagnetic N\'eel order, stripy order, spiral order, and zig-zag order (for an illustration of the ordering patterns, see e.g. Ref.~\onlinecite{Kimchi2011}). In our approach, the inclusion of quantum fluctuations in the FRG flow enables us to treat the system far beyond the classical limit. However, despite the quantum nature of our method, we clearly see that in a wide parameter regime the phase boundaries remain roughly the same as compared the to classical case. In order to exemplify this result, Fig.~\ref{Fig-phases} shows the phase diagram for $\alpha=0.5$ with $0\leq J_2, J_3\leq1$. This cut through the parameter space exhibits stripy, spiral and zigzag ordered phases with boundaries which almost agree with the classical one
 s. Only in the vicinity of the limits $\alpha=0$ or $\alpha=1$ quantum fluctuations have a more significant effect on the phase diagram and may even destroy the long-range order, see Fig.~3 of the main text. 

\paragraph*{Magnetic order.--}
The different types of order in the phase diagram of Fig.~\ref{Fig-phases} also reveal themselves through their different peak structures of the spin susceptibility in $\mathbf{k}$-space. Fig.~\ref{Fig-order} displays these susceptibility profiles in the entire extended (second) Brillouin zone for selected parameter points. In the case of stripy order, large response peaks can be seen at an $M$-point position (due to the external magnetic field along one of the cubic axes the susceptibility profile loses its six fold rotation symmetry). Upon entering the spiral regime, the ordering peak leaves the commensurate $M$ point and moves continuously towards the $\Gamma$ point. As one approaches the phase boundaries to the zig-zag phase, the ordering peaks broaden in $k_x$ direction and finally split into two separate peaks which mark the $\mathbf{k}$-space positions of the commensurate zig-zag order. These peaks are located in the middle between the $\Gamma$ point and the corners of
  the extended Brillouin zone. Recall that there are also three distinct zig-zag ordering configurations corresponding to three differently oriented zig-zag paths in the hexagonal lattice. Each of these configurations relates to two ordering peaks at $\mathbf{k}$ and $-\mathbf{k}$. As can be seen in the right plot of Fig.~\ref{Fig-order} the symmetry-breaking property of the external magnetic field only allows for two zig-zag ordering configurations, which corresponds to four peaks in the Brillouin zone.  

\paragraph*{Thermodynamic properties.--}
Finally, we turn to thermodynamic properties of model \eqref{Eq:NewHamiltonian}, in particular the suppression of magnetic ordering with regard to the Curie-Weiss temperature, which is often considered a measure of frustration and quantified by the frustration parameter $f=\theta/T_N$, i.e. the ratio of the Curie-Weiss temperature $\theta$ and the ordering temperature $T_N$. In Fig.~\ref{Fig-f}, we again consider the case $\alpha=0.5$. The qualitative behavior of $f$ shown there is characteristic also for smaller and larger values of $\alpha$. The stripy ordered phase can clearly be identified as the region with the smallest frustration, which is in agreement with the fact that this order is classically saturated at $\alpha=0.5$, $J_2=J_3=0$. As the boundary to the spiral phase is crossed, the frustration parameter undergoes a pronounced increase. In general, $f$ grows with increasing $J_2$ and $J_3$ (at least if $J_3$ is not too large) and the largest amount of frustration i
 s obtained near the transition line between spiral order and zig-zag order. Hence, by adding $J_2$ and $J_3$ interactions, the frustration can be enhanced by a factor of three as compared to the bare Heisenberg-Kitaev case. As shown in Fig.~3 of the main text, $f$ decreases monotonically with increasing $\alpha$, while the general profile of frustration in the $J_2$-$J_3$ plane remains similar to Fig.~\ref{Fig-f}. However, depending on $J_2$ and $J_3$, for small $\alpha$ a non-magnetic phase may appear, which causes a sharp rise of the frustration parameter in the nearby ordered phases. Hence, in regimes of small $\alpha$, the capability of $J_2$ and $J_3$ interactions to increase the frustration is most pronounced.

\end{widetext}

\end{document}